\documentclass[twocolumn,eqsecnum,aps]{revtex4} 
\def\Xint#1{\mathchoice
   {\XXint\displaystyle\textstyle{#1}}%
   {\XXint\textstyle\scriptstyle{#1}}%
   {\XXint\scriptstyle\scriptscriptstyle{#1}}%
   {\XXint\scriptscriptstyle\scriptscriptstyle{#1}}%
   \!\int}
\def\XXint#1#2#3{{\setbox0=\hbox{$#1{#2#3}{\int}$}
     \vcenter{\hbox{$#2#3$}}\kern-.5\wd0}}

\def\dashint{\Xint-}
\usepackage{epsfig}
\usepackage{psfrag}
\begin{document} 
\title{Revised Phase Diagram of the Gross-Neveu Model}  
\author{Michael Thies and Konrad Urlichs} 
\address{Institut f\"ur Theoretische Physik III\\ 
Universit\"at Erlangen-N\"urnberg\\ 
Staudtstra\ss e 7\\ 
D-91058 Erlangen\\ 
Germany} 
\date{\today} 
\begin{abstract}
We confirm earlier hints that the conventional phase diagram of the discrete chiral Gross-Neveu model 
in the large $N$ limit is deficient at non-zero chemical potential. We present the corrected phase diagram constructed  
in mean field theory. It has three different phases, including a kink-antikink crystal
phase. All transitions are second order. The driving mechanism for the new structure 
of baryonic matter in the Gross-Neveu model is an 
Overhauser type instability with gap formation at the Fermi surface.  
\end{abstract} 

\maketitle 

\section{Introduction} 

The simplest variant of the Gross-Neveu (GN) model is a 1+1 dimensional relativistic field theory
with $N$ species of fermions interacting via a quartic self-interaction \cite{R1},
\begin{equation}
{\cal L}= \bar{\psi} {\rm i}\gamma^{\mu} \partial_{\mu} \psi + \frac{1}{2} g^2 (\bar{\psi}\psi )^2 \ ,
\label{a1}
\end{equation}
where we suppress the flavour indices ($\bar{\psi}\psi \equiv \sum_i \bar{\psi}_i \psi_i$ etc.). 
This model possesses a discrete chiral symmetry ($\psi \to \gamma^5 \psi$)  which forbids a bare mass term.
In the large $N$ limit, it can be solved exactly. It exhibits a surprisingly large number of 
phenomena of interest to nuclear and particle physics (for a recent review, see Ref. \cite{R2}). Asymptotic freedom,
dimensional transmutation, spontaneous breakdown of 
chiral symmetry, dynamical mass generation, quark-antiquark and multiquark bound states are
its most prominent features. 
Moreover, the GN model is believed to have a non-trivial phase diagram. The discrete chiral symmetry gets
restored at high temperatures and/or densities. According to common lore, a critical line which
separates the chirally symmetric phase (massless fermions) from the broken phase (massive fermions)
comprises both first and second order lines separated by a tricritical point \cite{R3}. 
Models for which one can construct the renormalized phase diagram analytically
are extremely scarce and worth studying on theoretical grounds, even if they are far from being 
realistic. In the present case it is essential to define the model by letting $N \to \infty$ before taking
the thermodynamic limit. In this way one can avoid conflict with no-go theorems characteristic
for low dimensional systems \cite{R3a,R3b}. In particular, spontaneous breaking of continuous symmetries
becomes viable in 1+1 dimensions \cite{R4}. In spite of all the obvious reservations we feel that one can learn more about
relativistic matter from Eq. (\ref{a1}) than one had any right to hope on the basis of such an incredibly simple model. 

Given the large number of papers dealing with the GN model, one may wonder whether there are any
open questions left. This is actually the case. In the present work, we shall reconsider the model at finite
temperature and chemical potential. As we have pointed out recently, the alleged phase diagram
of the GN model is likely to be flawed \cite{R2}. It does not do justice to the role played by baryons in the
structure of baryonic matter through effects which are not $1/N$ suppressed.
Formally, the problem with previous investigations lies in
the tacit assumption of unbroken translational invariance. The condensate $\langle \bar{\psi}\psi \rangle$
then gives rise merely to an $x$-independent mass term. Following similar studies of the GN model 
with continuous chiral symmetry \cite{R5,R5a}, we have found evidence  
for a crystal phase with lower energy than the standard massive Fermi gas at zero temperature
 --- a kind of kink-antikink crystal \cite{R6}. This result was based on a variational calculation 
using a specific ansatz. In the present work, we report on a complete
Hartree-Fock calculation at finite $T$ and $\mu$. This enables us to round off our series of investigations by
presenting the revised phase diagram of the discrete chiral GN model in ``full glory".
The basic theoretical tool we have employed here is the relativistic Hartree-Fock method, appropriate for the 
leading order term of the large $N$ expansion. The same method has been used in
Refs. \cite{R6a}, \cite{R5} and \cite{R6} to which we refer the reader
for further background material. 

This paper is organized as follows: In Sect. II, we explain how we have solved the Dirac-Hartree-Fock equation
using a combination of numerical and analytical methods. Sect. III is dedicated to cold baryonic matter ($T=0$).
In Sect. IV, we collect the necessary formulae for finite temperature and chemical potential calculations.
Sect. V contains the main result of this work, namely the revised phase diagram of the GN model. In Sect. VI,
we summarize our findings, compare them with the continuous chiral GN model and 
put them into perspective.

\section{Spectrum of the single particle Dirac Hamiltonian}
 
We first discuss how to solve the Dirac-Hartree-Fock equation. In the GN model with
$( \bar{\psi}\psi )^2$-interaction, a mean field approach can only generate a (local) Lorentz scalar potential.
Therefore we have to diagonalize a single particle Dirac Hamiltonian of the generic form
\begin{equation}
H=H_0+V =\gamma^5 \frac{1}{\rm i} \frac{\partial}{\partial x} + \gamma^0 S(x)\ .
\label{d1}
\end{equation}
We choose the following representation of the $\gamma$-matrices,
\begin{equation}
\gamma^0 =-\sigma_1\ , \quad \gamma^1= {\rm i} \sigma_3 \ , \quad  \gamma^5 = - \sigma_2 \ .
\label{d1f}
\end{equation}
For spatially constant $S(x)$, $H$ reduces
to a free, massive Dirac Hamiltonian and our task would be rather trivial. This would evidently
lead back to the standard results for the GN model. Instead, we admit the possibility that
translational invariance breaks down and allow for spatially periodic $S(x)$ with 
(yet to be determined) period $a$,
\begin{equation}
S(x+a)=S(x) \ .
\label{d1a}
\end{equation}  
We first diagonalize the free Hamiltonian $H_0$. This step serves both to generate a basis for the
full, numerical diagonalization of $H$ and as a starting point for perturbation theory. 
Enclosing the system in a box of length $L=Na$ and 
imposing antiperiodic boundary conditions, we find
($\eta$ denotes the sign of the energy) 
\begin{equation}
H_0 |\eta,n\rangle  = \eta |k_n| |\eta ,n \rangle \qquad (\eta=\pm 1) \ , 
\label{d2}
\end{equation}
with the normalized spinor wave functions
\begin{equation}
\langle x| \eta ,n \rangle =\frac{1}{\sqrt{2L}}\left( \begin{array}{c} 1 \\ -{\rm i} \eta \,{\rm sgn}(k_n) \end{array}
\right){\rm e}^{{\rm i}k_n x}  ,  \ k_n = \frac{2\pi}{L} (n+1/2) .
\label{d3}
\end{equation}
Matrix elements of the potential $V=\gamma^0 S(x)$ in the unperturbed basis can be expressed via the
Fourier components
of $S(x)$
\begin{equation}
S(x)=\sum_{\ell} S_{\ell} {\rm e}^{{\rm i}2\pi \ell x/a}  \qquad (S_{-\ell} = S_{\ell}^*) 
\label{d5}
\end{equation}
as follows,
\begin{equation}
\langle \eta',n'|V|\eta,n \rangle = \frac{\rm i}{2} \left[ \eta \, {\rm sgn}(k_n)-\eta'{\rm sgn}(k_{n'})\right]
\sum_{\ell} \delta_{n'-n,N\ell} S_{\ell} .
\label{d6}
\end{equation}
Eqs. (\ref{d2}) and (\ref{d6}) are all that is needed for the numerical diagonalization of $H$. 
Due to the spatial periodicity of $S(x)$, the Hamiltonian matrix assumes a block diagonal form (conservation of
Bloch momentum). Further simplifications occur due to additional conservation laws for 
certain types of potential. Most of our calculations are based on the additional assumptions
\begin{eqnarray}
S(-x)&=&-S(x) \ , \nonumber \\
S(x+a/2) &=& -S(x) \ . 
\label{d13}
\end{eqnarray}
Such a scalar 
potential consists of a regular succession of positive and negative bumps of identical, symmetric shape.
For this choice, only odd $\ell$ appear, the $S_{\ell}$ are purely imaginary and the Hamiltonian matrix
is real symmetric, cf. Eq. (\ref{d6}). In some cases, we have employed more general shapes of the 
potential which then lead to larger, complex matrices. It turns out though that the self-consistent Hartree-Fock
potential belongs to the more restricted class of Eq. (\ref{d13}).

A characteristic difficulty of the relativistic approach is the fact that one has to determine the Dirac
sea self-consistently, not just the valence levels. In observables like the energy
density or the baryon density
infinities are then unavoidable. They require regularization and renormalization
and hence a certain amount of analytical work. Fortunately, due to asymptotic freedom of the GN model, fermion 
states with large momenta are only weakly affected by the mean field. Thus all UV-divergencies 
can be handled perturbatively. Perturbation theory has turned out to be very useful
to the present study  for other reasons as well. It provides more analytical insight into various aspects
of the phase diagram,
in particular the vicinity of the chirally restored phase where $S(x)$ is small.
Therefore, we briefly outline how to apply perturbation theory to the Hamiltonian
(\ref{d1}), restricting ourselves to potentials satisfying Eqs. (\ref{d13}).   

Naive application of 2nd order perturbation theory in $V$ would yield the following energies,
\begin{equation}
E_{\eta,n} \approx \eta \, {\rm sgn}(k_n) \left( k_n + \sum_{\ell} \frac{|S_{\ell}|^2}{2(k_n-q_{\ell})} \right) \ ,
\label{d7}
\end{equation}
with
\begin{equation}
q_{\ell} =  \frac{\pi \ell}{a}  \ .
\label{d8}
\end{equation}
Eq. (\ref{d7}) displays simple poles which signal the breakdown
of naive perturbation theory in the vicinity of the gaps. 
As is well known from solid state physics, this problem can be cured by resorting to ``almost degenerate
perturbation theory" (ADPT) in the region of the gaps \cite{R7}. More precisely, if $S_{\ell} \neq 0$
and we focus on the momentum region around $q_{\ell}$, we have to diagonalize $H$
exactly in two-dimensional subspaces of states whose momenta differ by a reciprocal lattice 
vector ($2 q_{\ell}$) but who have similar energy. For $\ell=0$, these are states with
opposite $\eta$ and the same (small) momentum $k_n$;
for $\ell \neq 0$, states with the same $\eta$ but momenta $k_n \pm q_{\ell}$, where
$k_n$ is again small. 
The result can be stated concisely as follows: Near $q_{\ell}$, replace the term which blows up in
Eq. (\ref{d7}) as follows,
\begin{equation}
\frac{|S_{\ell}|^2}{2(k_n-q_{\ell})} \longrightarrow {\rm sgn}(k_n-q_{\ell})\sqrt{(k_n-q_{\ell})^2+|S_{\ell}|^2}
-(k_n-q_{\ell})  .
\label{d9}
\end{equation}
If one seeks a reasonable ``global" approximation to the spectrum of $H$, one should apply Eq. (\ref{d9}) 
in the region between the midpoints of two gaps, i.e., 
\begin{equation}
k_n \in I_{\ell}=  \left[ \frac{q_{\ell}+q_{\ell-1}}{2},\frac{q_{\ell+1}+q_{\ell}}{2}\right] \ .
\label{d10}
\end{equation}
Hence the perturbative eigenvalues are best defined piecewise as
\begin{eqnarray}
E_{\eta,n}^{(\ell)}&\approx & \eta \, {\rm sgn}(k_n) \left( q_{\ell} + {\rm sgn}(k_n-q_{\ell}) \sqrt{(k_n-q_{\ell})^2+|S_{\ell}|^2}
\right. \nonumber \\
& & \left. +\sum_{j (\neq \ell)} \frac{|S_j|^2}{2(k_n-q_j)}\right)\  (k_n \in I_{\ell}) \ .
\label{d11}
\end{eqnarray}
As is well known, perturbatively each $S_{\ell} \neq 0$ induces exactly one gap at $k_n \approx q_{\ell}$
with width $2|S_{\ell}|$.

Our last remark concerns the standard solution of the GN model which corresponds
in our notation to $S_0 \neq 0$ only. In this particular case, ADPT
yields the exact spectrum, since the interaction only mixes the unperturbed states pairwise. 

\section{Baryonic matter at zero temperature}

We first present our results for the ground state of the GN model at finite baryon density 
($T=0$). We have to compute the Hartree-Fock energy density for a 
scalar potential $S(x)$,
\begin{equation}
\frac{{\cal E}_{\rm HF}}{N} = \frac{1}{L} \sum_{\eta,n}^{\rm occ} E_{\eta,n} + \frac{1}{2 Ng^2 L} \int_0^L
{\rm d}x S^2(x) \ .
\label{d14}
\end{equation}
For standard baryonic matter, the sum over
occupied states includes all negative energy states (the Dirac sea) and as many positive energy
states as required by the prescribed baryon density. Alternatively, one can consider ``antimatter"
where an appropriate number of negative energy states is unoccupied.
We adopt here the latter, more economical
way of evaluating the energy density.
The 2nd term in Eq. (\ref{d14}) corrects for double counting of the interaction energy in the sum 
over single particle energies. Since the self-consistent potential must be
of the form of a local, scalar potential in the GN model, it is sufficient to  
minimize the energy (\ref{d14}) with respect to $S(x)$ in order to find the full Hartree-Fock solution. 

Eq. (\ref{d14}) is not yet in a form suitable for numerical calculations.
The sum is badly UV divergent and the bare coupling constant appears in the 2nd term.
 Let us isolate the UV divergent terms
in the sum as follows: The sum over negative energy states with short wavelengths is
evaluated using perturbation theory. Provided we stay below the energy of the last
gap, we can simply use Eq. (\ref{d7}),
\begin{eqnarray}
\frac{2}{L} \sum_{n=\bar{n}}^{n_{\Lambda}} E_{-1,n} &\approx & -2 \int_{\bar{k}}^{\Lambda} \frac{{\rm d}k}{2\pi}
\left( k + \sum_{\ell} \frac{|S_{\ell}|^2}{2(k-q_{\ell})}\right) \nonumber \\
& \approx & \frac{\bar{k}^2-\Lambda^2}{2\pi} + 
\sum_{\ell} \frac{|S_{\ell}|^2}{2\pi} \ln  \frac{\bar{k}-q_{\ell}}{\Lambda}
\label{d14a}
\end{eqnarray} 
($\bar{k}=k_{\bar{n}}+\pi/L $). 
The quadratic divergence is irrelevant and can be eliminated by subtraction. The logarithmic divergence
is cancelled by a corresponding divergence in the double-counting correction as can be shown with the help of
the gap equation at zero density \cite{R6a} (we choose units such that the vacuum fermion mass $m_0$ is equal to 1),
\begin{equation}
\frac{1}{Ng^2} = 2 \int_0^{\Lambda} \frac{{\rm d}k}{2\pi} \frac{1}{\sqrt{k^2+1}} = \frac{1}{\pi}\ln (2 \Lambda) \ .
\label{d19a}
\end{equation}
The final expression for the regularized and renormalized ground state energy of a system with Fermi momentum
labelled by $n_{\rm F}$ is then
\begin{equation}
\frac{{\cal E}_{\rm HF}}{N} =\frac{2}{L} \sum_{n=n_{\rm F}}^{\bar{n}} E_{-1,n} + \frac{\bar{k}^2}{2\pi}
+ \frac{1}{2\pi} \sum_{\ell >0} |S_{\ell}|^2 \ln \left[ 4(\bar{k}^2-q_{\ell}^2)\right]  .
\label{t12}
\end{equation}
Before turning to the results, we briefly discuss two limiting cases where some analytical insight can be
reached: The high density and the low density limit.

In the {\em high density limit}, perturbation theory becomes applicable by virtue of asymptotic freedom.
The only self-consistent perturbative solution we could find is one where $S_{\pm 1}\neq 0$ only
and all orbits below the first gap are filled ($k_{\rm F}=q_1=\pi/a$). Using Eq. (\ref{d11}) and the real
variable
\begin{equation}
\tilde{S}_1 = {\rm i} S_1 \ ,
\label{d14aa}
\end{equation}
the self-consistent potential reads
\begin{equation}
S(x)=2 \tilde{S}_1 \sin(2 \pi x/a) \ .
\label{d15}
\end{equation}
Performing the renormalization along the lines described above leaves us with
\begin{equation}
\frac{{\cal E}_{\rm HF}}{N} = \frac{k_{\rm F}^2}{2\pi} + \frac{\tilde{S}_1^2}{4\pi}
\left[ \ln(16 k_{\rm F}^2 \tilde{S}_1^2) -1 \right] \ ,
\label{d16}
\end{equation}
where the first term is the result for a free Fermi gas of massless quarks. Minimization with
respect to $\tilde{S}_1$ yields the equation
\begin{equation}
\tilde{S}_1 \ln (4 k_f \tilde{S}_1) = 0
\label{d17}
\end{equation}
with the two solutions
\begin{equation}
\tilde{S}_1=0,\frac{1}{4 k_{\rm F}} \ .
\label{d18}
\end{equation}
The non-trivial solution is lower in energy by
\begin{equation}
\frac{{\cal E}_{\rm HF}}{N}-\left( \frac{{\cal E}_{\rm HF}}{N} \right)_0 = - \frac{1}{64 \pi k_{\rm F}^2} \ .
\label{d19}
\end{equation}
Evidently at $T=0$ the discrete chiral symmetry does not get restored at any 
finite density. 

In the {\em low density limit}, $S(x)$ is expected to be given by well separated,
equidistant kinks and antikinks. We therefore approximate $S(x)$ by
$\tanh x$ for $0<x<\Delta$ and 1 for $\Delta < x < a/4$, then continue symmetrically. 
Evaluating the Fourier coefficients $\tilde{S}_{\ell}$ for such a potential, we obtain 
(in the limit $\Delta\to \infty$)
\begin{equation}
\tilde{S}_{\ell} = \frac{2 k_{\rm F}}{\sinh (\pi \ell k_{\rm F})} \ .
\label{t13a}
\end{equation}
The Fourier
amplitudes satisfy some kind of scaling law, $\ell \tilde{S_{\ell}}$ being a ``universal function" of $\ell k_{\rm F}$.  
Roughly, $\tilde{S}_{\ell} \sim 1/\ell$ and the convergence of the Fourier expansion
of $S(x)$ becomes quite slow.

We now turn to the numerical results at $T=0$. We have carried out the above sketched diagonalization procedure.
The renormalized energy density was then minimized with respect to the period $a$ and the Fourier
coefficients $S_{\ell}$ of the
periodic scalar potential. We typically work in a box of length $L=1000$ (in units where the dynamical fermion mass
in the vacuum is 1), taking into account Fourier components in Eq. (\ref{d5}) up to $\ell=23$ if necessary.
As far as the period $a$ is concerned, we confirmed the results of  
Ref. \cite{R6}: The system invariably adjusts its period such that the Fermi surface coincides
with the bottom of the first gap. This is of course reminiscent of the Overhauser effect in condensed 
matter physics \cite{R8} (for recent discussions in the context of QCD, see Refs. \cite{A0,A1,A2,A3}). 
The spatially averaged baryon density per flavour is then given by
\begin{equation}
\frac{\rho}{N}= \frac{k_{\rm F}}{\pi} = \frac{1}{a} \ .
\label{t13q}
\end{equation}
\begin{figure}[t]
  \begin{psfrags}
    \begin{center}
      \psfrag{KFERMI}{$k_F$}
      \psfrag{TILDES1}{$\tilde{S}_1$}
      \psfrag{NIEDR}[lB]{low dens.}
      \psfrag{HOHED}[lB]{high dens.}
      \epsfig{file=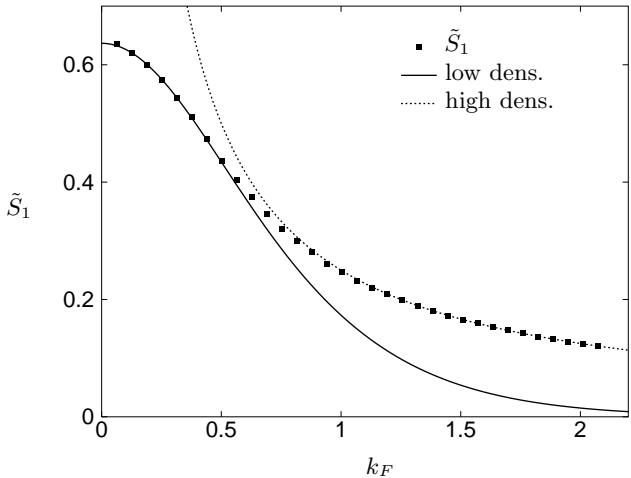, width=6cm, angle=-90}
    \end{center}
  \end{psfrags}
  \caption{Dominant Fourier component of scalar potential versus $k_{\rm F}$, compared with 
low- and high-density asymptotics.}
\end{figure}
In Fig. (1) we plot the most important Fourier component
($\tilde{S}_1$) as a function of $k_{\rm F}$ and compare it with the
two analytical limits discussed above. The numerically found Fourier
component interpolates nicely between the low and high density limits,
a good test of the diagonalization and minimization procedure.
Inclusion of higher Fourier components ($\ell=3,5,...$) is necessary
at lower densities. After minimization of the energy the corresponding
$S_{\ell}$ are practically indistinguishable from the analytical
result, Eq.  (\ref{t13a}), so that we refrain from showing any plot.
\begin{figure}[t]
  \begin{psfrags}
    \begin{center}
      \psfrag{latex-Sx}{$S(x)$}
      \psfrag{latex-x}{$x$}
      \psfrag{latex-013}[lB]{$k_F=0.13$}
      \psfrag{latex-025}[lB]{$k_F=0.25$}
      \psfrag{latex-101}[lB]{$k_F=1.01$}
      \psfrag{latex-251}[lB]{$k_F=2.51$}
      \epsfig{file=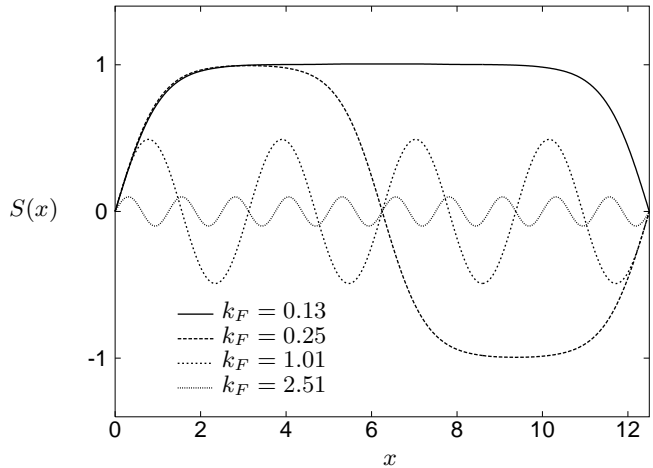, width=6cm, angle=-90}
    \end{center}
  \end{psfrags}
  \caption{Spatial dependence of Hartree-Fock potential at various densities.}
\end{figure}
A representative selection of the self-consistent Hartree- Fock potentials $S(x)$ is shown in Fig. (2) for a range
of densities. This figure illustrates how $S(x)$ interpolates between well separated kinks and antikinks
($\sim \tanh x$) at low density and the perturbative potential ($\sim  \sin 2 k_{\rm F}x$) at high density.
\begin{figure}[t]
  \begin{psfrags}
    \begin{center}
      \psfrag{EPSILONHF}{$\displaystyle \frac{\cal E_{\rm HF}}{N}$}
      \psfrag{KFERMI}{$k_F$}
      \psfrag{APPROX}{\parbox[t]{2cm}{\raggedright analytical \\ approximation}}
      \psfrag{MATCHING}{matching}
      \epsfig{file=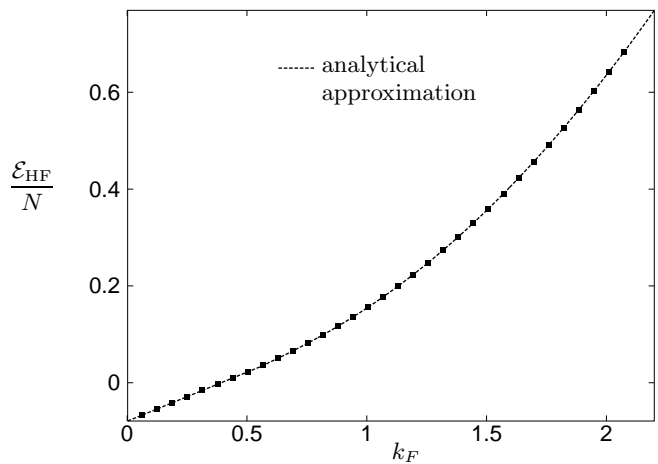, width=6cm, angle=-90}
    \end{center}
  \end{psfrags}
  \caption{Hartree-Fock energy density compared with analytical approximation
obtained by asymptotic matching.}
\end{figure}

The energy density obtained from our numerical calculation 
is displayed in Fig. (3). The dashed curve is an analytical approximation obtained by 
matching the asymptotic behaviour at large and small $k_{\rm F}$ as follows: For low
density, we have widely separated baryons of mass $2N/\pi$, therefore
\begin{equation}
\frac{{\cal E}_{\rm HF}}{N} \approx  -\frac{1}{4\pi}+ \frac{2 k_{\rm F}}{\pi^2} \qquad (k_{\rm F} \to 0) \ . 
\label{t13h}
\end{equation} 
At high density, perturbation theory predicts [cf. Eqs. (\ref{d16}) and (\ref{d19})]
\begin{equation}
\frac{{\cal E}_{\rm HF}}{N} \approx  \frac{k_{\rm F}^2}{2\pi} - \frac{1}{64 \pi k_{\rm F}^2} \qquad (k_{\rm F}\to \infty)\ . 
\label{t13i}
\end{equation}
The two expressions (\ref{t13h}) and (\ref{t13i}) have been matched at the point where they agree ($k_{\rm F}=0.55537$).
This simple estimate seems to catch the essence of a lengthy numerical calculation amazingly well.

\begin{figure}[t]
  \begin{psfrags}
    \begin{center}
      \psfrag{EPSILONHF}{$\displaystyle \frac{\cal E_{\rm HF}}{N} - \left(\frac{\cal E_{\rm HF}}{N}\right)_{\mathrm{0}}$}
      \psfrag{KFERMI}{$k_F$}
      \psfrag{APPROX}{\parbox[t]{2cm}{\raggedright perturbative \\ prediction}}
      \psfrag{DELTAE}{$\Delta \cal E$}
      \psfrag{MATCHING}{matching}
      \epsfig{file=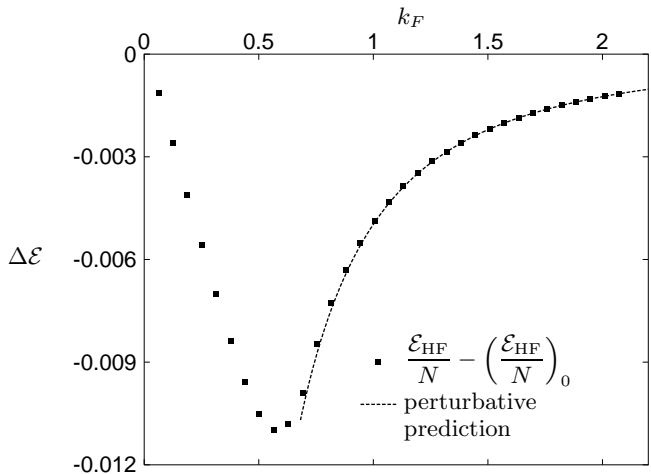, width=6cm, angle=-90}
    \end{center}
  \end{psfrags}
  \caption{Ground state energy difference between kink-antikink crystal and translationally invariant solution.}
\end{figure}

In Fig. (4), we show the energy difference between our solution and the standard solution of the
GN model based on unbroken translational invariance (the dashed line is the perturbative 
prediction).
The kink-antikink crystal is energetically favoured at all densities. Our results are of
course also lower than the more restricted variational calculation of the kink-antikink crystal in Ref. \cite{R6},
the improvement being most significant at high densities.
Finally, we can derive the baryon chemical potential from the slope of the curve in Fig. (3),  
\begin{equation}
\mu = \frac{\partial {\cal E}_{\rm HF}}{\partial \rho} \ .
\label{t13j}
\end{equation} 
\begin{figure}[t]
  \begin{psfrags}
    \begin{center}
      \psfrag{MU}{$\mu$}
      \psfrag{RHON}{$\displaystyle \frac{\rho}{N}$}
      \psfrag{APPROX}{Fermi gas}
      \psfrag{EINSPI}{$\frac{2}{\pi}$}
      \psfrag{EINSWURZEL}{$\frac{1}{\sqrt{2}}$}
      \psfrag{CRYSTAL}{crystal}
      \psfrag{MUPI}{$\displaystyle \sim\frac{\mu}{\pi}$}
      \epsfig{file=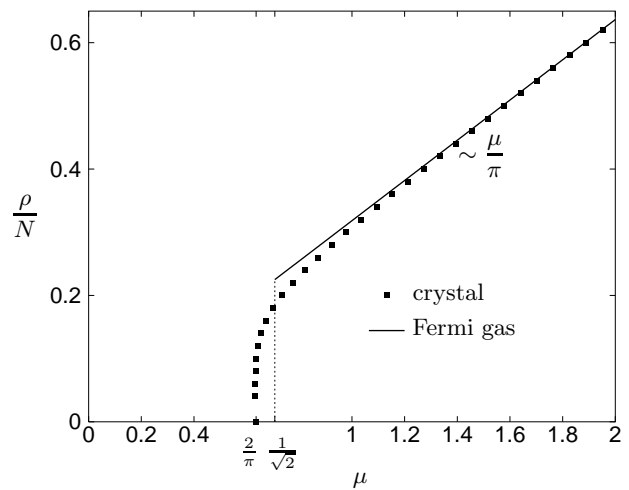, width=6cm, angle=-90}
    \end{center}
  \end{psfrags}
  \caption{Mean baryon density versus chemical potential at $T=0$, compared to the translationally invariant case.}
\end{figure}
In Fig. (5) we have plotted $\rho/N$ vs. $\mu$. As expected on the basis of the low density limit, the curve
starts at $M_B/N=2/\pi$. It then rises steeply and approaches an asymptote.
The solid curve is the result of the conventional calculation which involves a Maxwell-type construction.
Here, the density jumps discontinuously at $\mu = 1/\sqrt{2}$ where a first order phase transition takes place.
Our calculation is strongly in favour of a 2nd order phase transition at $\mu=2/\pi$ (or $\rho=0$). 
These results at $T=0$ already indicate that the commonly accepted phase diagram of the GN model is not tenable.

\section{Grand canonical potential}

In this section, we collect the formulae needed for determining the phase diagram and computing
thermodynamic observables in the grand canonical framework (see e.g. \cite{R9}). 
Starting point for the full, non-perturbative calculation is the formal expression for the grand
canonical potential density $\Psi$ ($\beta=1/T$), 
\begin{eqnarray}
\frac{\Psi}{N} &=& -\frac{2}{\beta L} \sum_{n=0}^{n_{\Lambda}} \ln \left[ \left(1+{\rm e}^{-\beta (E_n-\mu)}\right)
\left( 1+{\rm e}^{\beta (E_n+\mu)}\right)\right] \ \nonumber \\
&+&
\frac{1}{2 Ng^2 L} \int_0^L {\rm d}x S^2(x) \ .
\label{t6}
\end{eqnarray}
This quantity has to be minimized with respect to $S(x)$.
From now on, single particle energies will be defined as positive energies, i.e., $E_n \equiv E_{+1,n}$.
The sums over positive and negative energy states correspond to the two
factors under the logarithm. 
The first term in Eq. (\ref{t6})
is easily recognized as the ``effective potential" expected for independent fermions. The 2nd term is identical 
to the double counting correction at zero temperature. Eq. (\ref{t6}) is still symbolic
due to UV divergencies and requires some modifications.
We proceed exactly like in the $T=0$ case, dividing up   
the spectrum into two parts: For $n=0...\bar{n}$, the energies $E_n$ are obtained by diagonalization
of the Dirac-Hartree-Fock Hamiltonian in a large but finite box. For
$n>\bar{n}$ we use the perturbative single particle energies and replace the sum by an integral. 
Using the notation
\begin{equation}
\varepsilon (k) = k + \sum_{\ell >0} \frac{|S_{\ell}|^2}{2} \left( \frac{1}{k-q_{\ell}}+ \frac{1}{k+q_{\ell}}\right)
\label{t7}
\end{equation}
for the perturbative energies and $\bar{k}$ as defined after Eq. (\ref{d14a}), we find
\begin{eqnarray}
\frac{\Psi}{N} &=&\  - \frac{2}{\beta L} \sum_{n=0}^{\bar{n}} \ln \left[ \left(1+{\rm e}^{-\beta (E_n-\mu)}\right)
\left( 1+{\rm e}^{\beta (E_n+\mu)}\right)\right]  \nonumber \\
&-&  \frac{2}{\beta} \int_{\bar{k}}^{\infty} \frac{{\rm d}k}{2\pi} \ln \left[ \left(1+{\rm e}^{-\beta (\varepsilon(k)-\mu)}\right)
\left( 1+{\rm e}^{-\beta (\varepsilon(k)+\mu)}\right)\right] \nonumber \\
&+&  \frac{\bar{k}^2}{2\pi} + \frac{\mu \bar{k}}{\pi} + \frac{1}{2\pi}
\sum_{\ell >0} |S_{\ell}|^2 \ln \left[ 4(\bar{k}^2-q_{\ell}^2)\right] \ .
\label{t8}
\end{eqnarray}
Note carefully the signs in the various exponents. They arise because one has to pull out the sum over single particle
energies for $k>\bar{k}$ in order to perform the renormalization, cf. Sect. III.
As usual we have dropped the irrelevant divergent terms
\begin{equation}
- \frac{\Lambda^2}{2\pi} - \frac{\Lambda \mu}{\pi}\ .
\label{t8a}
\end{equation}
The quadratically divergent term is already familiar from $T=0$, the linearly divergent term reflects 
the infinite baryon density
of the Dirac sea. From Eq. (\ref{t8}), all other thermodynamic quantities (notably pressure $P$, baryon density $\rho$,
entropy density $s$, and energy density $u$) follow
in the standard way,
\begin{eqnarray}
P & = &  - \Psi \ , \qquad \qquad \rho \ =\  - \frac{\partial}{\partial \mu} \Psi \ , \nonumber \\
s & = & \beta^2 \frac{\partial}{\partial \beta} \Psi \ , \qquad \  u \ =\  Ts-P+\mu \rho \ . 
\end{eqnarray}
In the chirally restored phase ($S(x)=0$), everything 
can be trivially evaluated in closed form,
\begin{eqnarray}
\frac{P}{N} &=& \frac{\pi}{6} T^2 + \frac{\mu^2}{2\pi} - \frac{1}{4\pi}\ , \qquad
\frac{\rho}{N} \ =\ \frac{\mu}{\pi}\ , \nonumber \\
\frac{u}{N} &=& \frac{\pi}{6} T^2 + \frac{\mu^2}{2\pi} + \frac{1}{4\pi} ,\qquad \frac{s}{N} \ =\ \frac{\pi}{3} T\ . 
\label{t30a}
\end{eqnarray}

\section{Revised phase diagram}

According to common lore, the GN model has two
phases \cite{R3}: A chirally broken phase with a dynamical fermion mass  $m\neq 0$ at
low ($\mu, T$), and a chirally symmetric phase with massless fermions at high $\mu$ and/or $T$
(chiral symmetry referring to
the discrete symmetry $\psi \to \gamma^5 \psi$). These two phases are supposedly separated
by a 2nd order line going from ($\mu=0,T=T_c$)
to ($\mu=\mu_t, T=T_t$) and a 1st order line from the latter point to ($\mu=1/\sqrt{2}, T=0$).
The critical temperature at zero chemical potential has the value $T_c={\rm e}^{\rm C}/\pi = 0.56693$, whereas the tricritical 
point is located at $\mu_t=0.60822 , T_t=0.31833$. A mixed phase also appears which 
can be mapped out in other types of phase diagrams like the ($u,\rho$)- or ($P, 1/\rho$)-plots, cf. Ref. \cite{R10}. 

At variance with these results we have clearly identified three distinct phases in the ($\mu,T$)-diagram. In addition to the two
known ones there is a crystalline phase with broken chiral and translational symmetry. 
These three phases are all separated by 2nd order lines which meet in one point.

We first explain how we have obtained the boundary of the chirally restored phase.
A 2nd order line separates the chirally restored phase from either the massive
or the crystal phase, depending on $\mu$. Since the scalar potential vanishes continuously 
across this line, the line itself can be determined rigorously by using perturbation theory.
We start from Eq. (\ref{t6}), add counter terms to the linearly and quadratically divergent
pieces of Eq. (\ref{t8a}) and take the limit $L\to \infty$, replacing the sum by an integral.
Setting $E(k)=k+\Delta(k)$ and linearizing in $\Delta(k)$ yields
\begin{eqnarray}
\frac{\Psi}{N} & = & -\frac{2}{\beta}
\int_0^{\Lambda} \frac{{\rm d}k}{2\pi} \ln \left[ (1+{\rm e}^{-\beta(k-\mu)})
(1+{\rm e}^{\beta(k+\mu)})\right]
\nonumber \\
 &+&  \frac{\Lambda^2}{2\pi} + \frac{\Lambda \mu}{\pi} + \frac{1}{2\pi}  \ln (2\Lambda)\sum_j |S_j|^2 
\nonumber \\
&-& 2 \int_0^{\Lambda} \frac{{\rm d}k}{2\pi}
\Delta(k) \frac{\sinh \beta k}{\cosh \beta k + \cosh \beta \mu}  \ .
\label{t9}
\end{eqnarray}
Let us now assume that translational symmetry is also broken spontaneously and that 
\ --- just like in Sect. III at zero temperature ---\ 
only $S_{\pm 1} \neq 0$. Using the results for ADPT of Sect. II,
we find (with $q \equiv q_1=\pi/a$)  
\begin{eqnarray}
\Delta(k) &=& q -k + {\rm sgn}(k-q)\sqrt{(k-q)^2+|S_1|^2} + \frac{|S_1|^2}{2(k+q)} 
\nonumber \\
& & \qquad \qquad \qquad  {\rm for} \ \ k \in \left[\frac{q}{2},
\frac{3q}{2}\right] \ ,
\nonumber \\
\Delta(k) & = &  \frac{|S_1|^2}{2} \left( \frac{1}{k-q} + \frac{1}{k+q}\right)
\nonumber \\
& & \qquad \qquad \qquad  {\rm for} \ 0<  k<q/2 \ , k > 3 q/2 \ .
\label{t14}
\end{eqnarray}
We can actually expand the square root in $|S_1|^2$ provided we are careful about the treatment of the resulting pole
in the integrand.
One can easily convince oneself that one has to take the Cauchy principal value part
(this being the only remnant of ADPT). Dropping those
terms which do not depend on $S_1$ then yields
\begin{eqnarray}
\frac{\Psi}{N}&\approx & \frac{1}{\pi} |S_1|^2  \Biggl[ \ln (2\Lambda) \Biggr. \label{170} \\
&-& \left.  \dashint_0^{\Lambda}  \ {\rm d}k
\left( \frac{k}{k^2-q^2}\right)\frac{\sinh \beta k}{\cosh \beta k+\cosh \beta \mu}\right] \ .
\nonumber 
\end{eqnarray}
The phase boundary for the 2nd order transition under the constraint that $q$ (or $a$) has a given value 
is obtained from the condition that the r.h.s. of Eq. (\ref{170}) vanishes. 
For each $q$ this defines a curve in the ($T,\mu$)-plane. After a rescaling of variables,
\begin{equation}
\kappa= \beta k \ , \quad \mu' = \beta \mu\ , \quad \Lambda' = \beta \Lambda \ ,\quad q'=\beta q \ ,
\label{z17}
\end{equation}
we can solve explicitly the condition that the coefficient of $|S_1|^2$ in Eq. (\ref{170}) vanishes with the result
\begin{equation}
\beta_{\rm crit}(\mu', q') = \lim_{\Lambda' \to \infty} 2 \Lambda' \exp \Phi(\mu',q',\Lambda') \ ,
\label{t18a}
\end{equation}
where
\begin{equation}
\Phi(\mu',q',\Lambda')=
  \dashint_0^{\Lambda'}  \ 
{\rm d}\kappa
\left( \frac{\kappa}{(q')^2 - \kappa^2 } \right)
\frac{\sinh \kappa}{\cosh \kappa + \cosh \mu'}  .
\label{t18}
\end{equation}
If we plot $T_{\rm crit}=1/\beta_{\rm crit}(\mu',q')$ against $\mu_{\rm crit}=\mu'/\beta_{\rm crit}$ for all possible values
of $q'$, the envelope of the resulting family of curves represents the sought for phase boundary.
Incidentally, since we allow $q$ to vary freely, we can also consider the limit $q \to 0$ in which 
the period of the crystal becomes infinite. In this limit one recovers the results of the standard GN
model solution which are usually derived by assuming $S_0 \neq 0$. Thus our method enables us to deal with the 
translationally broken and unbroken situation on the same footing.

\begin{figure}[t]
  \begin{psfrags}
    \begin{center}
      \psfrag{PUNKTA}{A}
      \psfrag{PUNKTB}{B}
      \psfrag{PUNKTC}{C}
      \psfrag{MU}{$\mu$}
      \psfrag{TEMPERATUR}{$T$}
      \psfrag{TRIPELPUNKT}{\parbox{2cm}{$T_t \approx 0.318$\\ $\mu_t\approx 0.608$}}
      \psfrag{TKRITISCH}{$T_\mathrm{crit}$}
      \epsfig{file=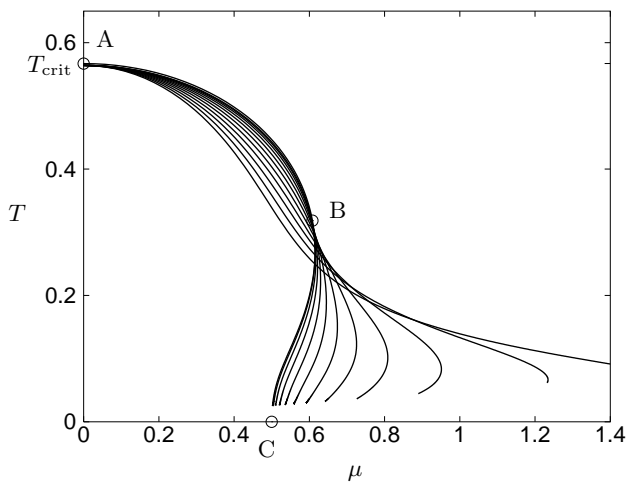, width=6cm, angle=-90}
    \end{center}
  \end{psfrags}
  \caption{Perturbative determination of phase boundaries as envelope of a family of curves.
Each individual curve correponds to a certain wave number of the periodic potential.}
\end{figure}

In Fig. 6, we display a few representative curves obtained from 
Eq. (\ref{t18}). In particular the curve $ABC$ corresponds to $q'=0$.
The section $AB$ coincides with the
2nd order line of the standard solution, $B$ being the tricritical point \cite{R3}. 
This part of the phase boundary survives if we relax the assumption of unbroken translational invariance.

The true phase boundary beyond the tricritical point can be generated
as envelope of the curves shown in Fig. (6). For each $\mu>\mu_t$ there is one particular curve (labelled by $q$)
which touches the envelope at this $\mu$-value. This allows us to define a function 
$q(\mu)$ which describes how the period
of the crystal depends on the chemical potential along the phase boundary.
This is of some interest since $q=\pi/a$ is an order parameter for the breakdown of translational
invariance. In Fig. (7) we show the dependence of the order parameter $q$ on $\mu$ as one moves along
the phase boundary.
The solid line is the curve $q=\mu$ which is approached asymptotically by the full calculation.  
At $\mu=\mu_t$, the tricritical point of the old solution, we see a clear signal of a 2nd order phase transition with breakdown
of translational invariance. The point $B$ in Fig. (6) is therefore also a multicritical point in the 
revised phase diagram. It plays a somewhat different role though. In fact, it has all the characteristics of a
``Lifschitz point" in condensed matter theory \cite{R11}. This type of multicritical point has been discussed in
 the context of magnetic or liquid crystals which exhibit both periodic and homogeneous ordered phases.
 At the Lifschitz point the wave vector of the periodic structure vanishes continuously, just like in Fig. (7).

\begin{figure}[t]
  \begin{psfrags}
    \begin{center}
      \psfrag{MU}{$\mu$}
      \psfrag{MUT}{$\mu_t$}
      \psfrag{WELLENVEKTOR}{$q$}
      \psfrag{QMU}{numerical results}
      \psfrag{QGLEICHMU}{$q=\mu$}
      \epsfig{file=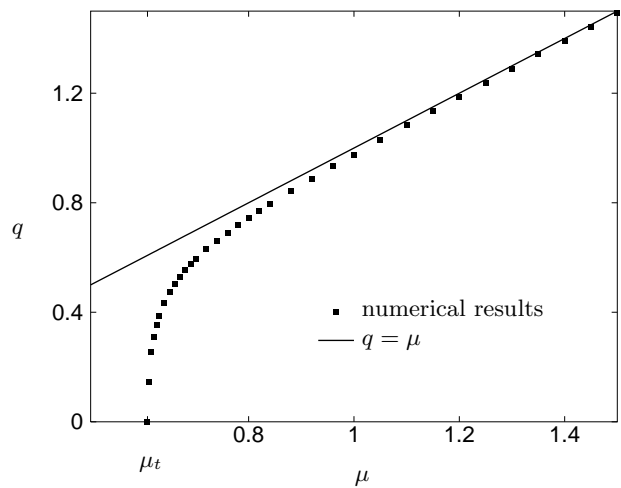, width=6cm, angle=-90}
    \end{center}
  \end{psfrags}
  \caption{Wavenumber characterizing the crystal period along the phase boundary, showing 
the presence of a ``Lifschitz point" at $\mu_t$.}
\end{figure}

One can determine the envelope of the curves from Fig. (6) numerically and display it in the ($T,\mu$) diagram.
The result is drawn in Fig. (8) up to $\mu=1.5$. For
larger values of the chemical potential, it is possible to derive the form of the phase boundary
analytically (see Appendix) with the result
\begin{equation}
T_{\rm crit} = \frac{{\rm e}^{\rm C}}{4\pi \mu} \ .
\label{in1}
\end{equation}
This curve does not intersect the $T=0$ axis, as expected on the basis of our $T=0$ results of Sect. III.
\begin{figure}[t]
  \begin{psfrags}
    \begin{center}
      \psfrag{MU}{$\mu$}
      \psfrag{ZWEIPI}{$\frac{2}{\pi}$}
      \psfrag{TEMPERATUR}{$T$}
      \psfrag{PUNKTA}{A}
      \psfrag{PUNKTB}{B}
      \psfrag{TKRITISCH}{$T_\mathrm{crit}$}
      \psfrag{WURZEL}{$\frac{1}{\sqrt{2}}$}
      \psfrag{CHIRAL}{\parbox{1.5cm}{\raggedright chirally \\ symmetric \\ ($m=0$)}}
      \psfrag{TRANSINV}{\parbox{1.3cm}{\raggedright translationally invariant ($m\neq 0$)}}
      \psfrag{CRYSTAL}{crystal phase}
      \epsfig{file=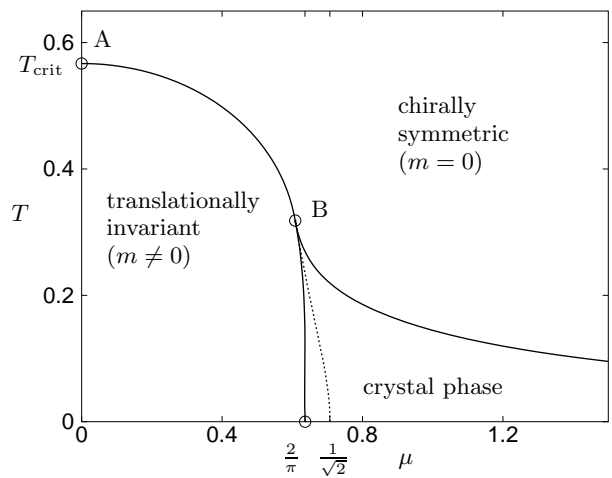, width=6cm, angle=-90}
    \end{center}
  \end{psfrags}
  \caption{Revised phase diagram of the GN model in the ($T,\mu$)-plane. The dashed line is the first order
 phase boundary from the previous, translationally invariant calculation.}
\end{figure}

From Fig. (5) at $T=0$ and Fig. (7) along the perturbative phase boundary one would expect
another 2nd order line joining the Lifschitz point ($T_t,\mu_t$) with the point ($T=0,\mu=2/\pi$). The scalar
potential is constant to the left of this hypothetical line and periodic to the right, its period $a$ diverging as one 
approaches the line from the right. Since the amplitude is non-vanishing on both sides, we can no longer invoke 
perturbation theory. 
We have determined this phase boundary numerically by computing the value of the 
grand canonical potential under the constraint that there are 0,2,4 potential wells in the interval of length $L$.
This is done along curves  of constant $T$, varying $\mu$ in the relevant region. At the critical line, all three values
of the potential
should converge for sufficiently large $L$. Thereby, one can locate the value of $\mu$ where the instability
with respect to breakdown of translational invariance (kink-antikink formation) sets in.
For technical reasons, this particular calculation was done in a much smaller interval ($L\approx 40-80$) but
the results are stable with respect to increasing $L$ within the accuracy of our plot. The resulting critical line
is also included in Fig. (8).
The tricritical point of the old solution coincides with the point where all three phase boundaries meet. 
The dashed curve is the first order line of the old solution \cite{R3} which has now become obsolete. It has only been included
to highlight in which way the phase diagram changes.
 
A perhaps physically more illuminating way of presenting the phase diagram is given in Fig. (9)
where we have transformed our phase boundaries into the ($T,\rho$)-plane. 
\begin{figure}[t]
  \begin{psfrags}
    \begin{center}
      \psfrag{RHO}{$\rho$}
      \psfrag{TEMPERATUR}{$T$}
      \psfrag{TKRITISCH}{$T_\mathrm{crit}$}
      \psfrag{CHIRAL}{\parbox{1.5cm}{\raggedright chirally \\ symmetric \\ ($m=0$)}}
      \psfrag{TRANSINV}{\parbox{1.3cm}{\raggedright translationally invariant ($m\neq 0$)}}
      \psfrag{CRYSTAL}{\ crystal \hspace{0.2mm} phase}
      \epsfig{file=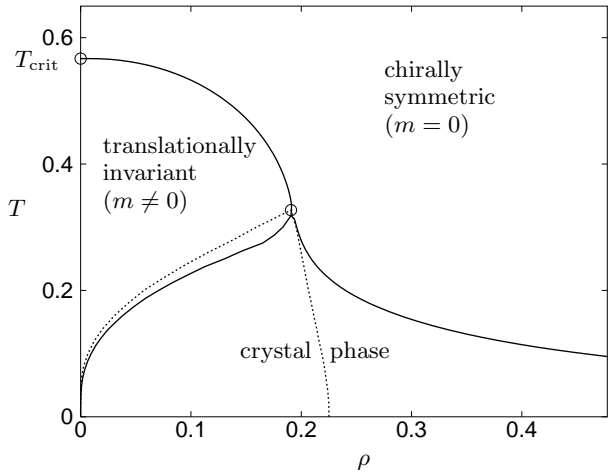, width=6cm, angle=-90}
    \end{center}
  \end{psfrags}
  \caption{Revised phase diagram of the GN model in the ($T,\rho$)-plane. The dashed lines 
belong to the old phase diagram, where they enclose the mixed phase. This ``droplet" region
is superseded by the crystal phase featuring baryons.}
\end{figure}
Here, the first order line of the old solution splits up into the two dashed lines which delimit the mixed phase
region (droplets of chirally restored matter in the chirally broken vacuum). This
should be replaced now by the two solid lines going downward from the former tricritical point and
enclosing the crystal phase. In the new solution at $T=0$ the crystal phase is stable at all finite densities. If one 
increases the temperature at fixed density, the outcome depends on the value of $\rho$. For $\rho>\rho_t=\mu_t/\pi$,
when crossing the new phase boundary one goes directly into the chirally- and translationally restored
phase in a second order transition. For $\rho < \rho_t$, translational invariance is restored first. At some
higher temperature the dynamical fermion mass vanishes and chiral symmetry gets restored as well.

Our results suggest that the phase transition between crystal and massive Fermi gas is also
a 2nd order transition. Since the corresponding phase boundary has only been obtained numerically,
we have computed various thermodynamic observables along isotherms in the ($T,\mu$)-plane
to check whether they are indeed continuous.
In the old solution, quantities like $\rho, s, u$ are discontinuous across the first order line. By contrast, we
see no discontinuity in any of these quantities within our numerical accuracy. This is illustrated
by way of example in Fig. (10) for
the entropy density $s$ as a function of $\mu$  (at $T=0.1$).
\begin{figure}[t]
  \begin{psfrags}
    \begin{center}
      \psfrag{MU}{$\mu$}
      \psfrag{ENTROPIE}{$s$}
      \psfrag{ENTROPDENS}{crystal}
      \psfrag{TRANSINV}{Fermi gas}
      \epsfig{file=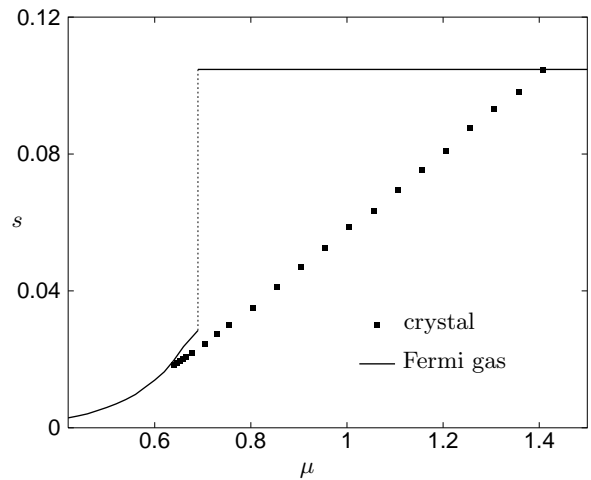, width=6cm, angle=-90}
    \end{center}
  \end{psfrags}
  \caption{Entropy density versus chemical potential at $T=0.1$ showing the absence of latent heat, 
in contrast to the translationally invariant case.}
\end{figure}
Whereas in the old solution the entropy exhibits a huge discontinuity
and jumps to the chirally restored value of Eq. (\ref{t30a}) at the first order phase transition, the crystal
solution interpolates smoothly between the massive and massless Fermi gas. It shows no sign whatsoever  of
a latent heat. The corresponding curves for $\rho$ and $u$ also favour a second order transition. Since here the variation
near the phase transition is more rapid and it is difficult to do accurate computations very close to the phase boundary, 
these other observables are less well suited for judging the order of the phase transition and will not be shown here.

Finally, we wish to point out that one can do more analytical work on the crystal phase in the asymptotic region
of large $\mu$ and large $\beta$. As derived in the appendix,
the relation
between $4 \mu \tilde{S}_1$ and $4 \mu T$ in the crystal phase at large $\mu$ is the same as the one
between $m$ and $T$ in the massive Fermi gas phase at $\mu=0$, see Fig. (11).
We do not really understand the reason for this interesting kind of scaling behaviour. It may point to 
additional simplifying features which we have not yet fully exploited.
\begin{figure}[t]
  \begin{psfrags}
    \begin{center}
      \psfrag{TEMPERATUR}{$4\mu T$}
      \psfrag{MASSE}{$4\mu \tilde{S}_1$}
      \epsfig{file=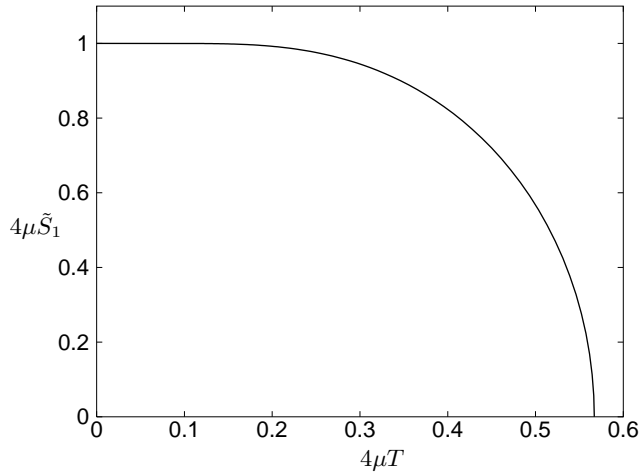, width=6cm, angle=-90}
    \end{center}
  \end{psfrags}
  \caption{Asymptotic behaviour of the amplitude of the periodic scalar potential versus temperature, valid for large $\mu$.
The curve shown is identical to the dependence of the fermion mass $m$ on $T$ at  $\mu=0$.}
\end{figure}

\section{Summary and perspective}

Let us summarize our findings about the phase structure of the GN model and
try to understand the essential underlying physics. Together with our earlier results 
for the GN model with continuous chiral symmetry \cite{R5}  (or equivalently the ``two-dimensional
Nambu--Jona-Lasinio (NJL$_2$) model", we are now in a position to cover both variants of the 
model and compare their phase structure. So far, it has been assumed that the phase diagrams
are the same for the discrete and continuous cases \cite{R10} and exhibit two phases: Massive and
massless fermions. The only symmetry issue considered was the breakdown of chiral symmetry.
The phase boundary in the ($\mu,T$)-diagram had lines of first and second order transitions
as well as a tricritical point. Due to the first order transition a mixed phase could also appear.

The central result of our investigation is the fact that both of these model theories can break
translational invariance as well. Since this is a continuous symmetry, the large $N$ limit is of 
course instrumental for the phase structure \cite{R4,R2}. We will now argue that the 
dynamical mechanism behind the breakdown of chiral symmetry in the vacuum and translational symmetry
in baryonic matter is actually the same. 

\begin{figure}[t]
  \begin{psfrags}
    \begin{center}
      \psfrag{DYNAM}{NJL$_2$ model}
      \psfrag{FREE}{free spectrum}
      \psfrag{WELLENVEKTOR}{$k/q$}
      \psfrag{EPSILON}{$E$}
      \epsfig{file=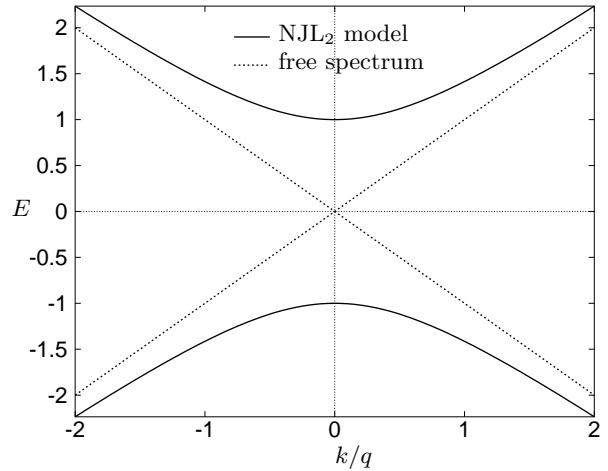, width=6cm, angle=-90}
    \end{center}
  \end{psfrags}
  \caption{Single particle energy spectrum for the NJL$_2$ model  with dynamical mass gap.}
\end{figure}

Our point can most clearly be made at $T=0$. Let us first recall the structure of baryonic matter
in the NJL$_2$ model. The condensate is best pictured as a helix in the space spanned by
$\langle \bar{\psi}\psi \rangle$, $\langle \bar{\psi} {\rm i}\gamma^5 \psi \rangle$, and $x$
(the ``chiral spiral" \cite{R5}). Since the chiral winding number is equal to baryon number, each full turn of the
helix correponds to one baryon.

 From the point of view of Hartree-Fock theory, it is instructive
to take a look at the single fermion spectrum. In the vacuum one has dynamical
mass generation with the spectrum sketched in Fig. (12). Due to the peculiar properties of this model, 
the whole picture drawn in Fig. (12) just moves upward or downward by an amount $\mu$
at finite chemical potential. In Fig. (13), we illustrate the filling of single particle orbits for vacuum, matter
and antimatter (at $T=0$), respectively. We see that there is always a gap of the same width ``floating" 
at the Fermi surface. This enables the system
to make optimal use of the level repulsion 
at the gap for lowering its energy.
\begin{figure}[t]
  \begin{psfrags}
    \begin{center}
      \psfrag{MU0}{$\mu=0$}
      \psfrag{MUG0}{$\mu > 0$}
      \psfrag{MUK0}{$\mu < 0$}
      \epsfig{file=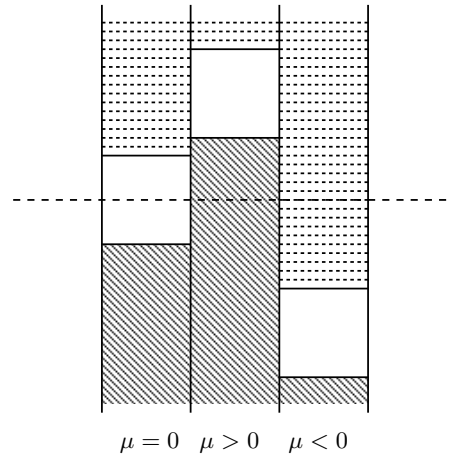, width=6cm}
    \end{center}
  \end{psfrags}
  \caption{Schematic plot illustrating the filling of single particle states in the NJL$_2$ model for vacuum
($\mu=0$), matter ($\mu>0$) and
antimatter ($\mu<0$), respectively. Dark dashed regions: filled orbits, light dashed: empty orbits, white: energy gaps.}
\end{figure}
The important dynamics is gap formation at the Fermi surface, which reduces to the surface of the
Dirac sea in the special case $\rho=0$. The system breaks whatever symmetry it takes to generate
the gap, either chiral symmetry in the vacuum or chiral and translational symmetry in matter
(with one unbroken combination of the two). This scenario has a lot in common with the Overhauser effect,
originally discovered in non-relativistic Hartree-Fock systems \cite{R8}. The nice feature about the present relativistic
application is the intimate relationship between  mass gap and band gap. Note that in the NJL$_2$-model
the single particle spectra for matter ($\mu>0$) and antimatter ($\mu<0$) are different, although
the energy densities agree (for the same $|\mu|$). This follows from the fact that under
charge conjugation
\begin{equation}
\psi_c(x)=\gamma^1 \psi^*(x) \ ,
\label{in10}
\end{equation}
the scalar potential is invariant but the pseudoscalar one changes sign.
\begin{figure}[t]
  \begin{psfrags}
    \begin{center}
      \psfrag{WELLENVEKTOR}{$k/q$}
      \psfrag{EPSILON}{$E$}
      \psfrag{CHIRAL}{GN model}
      \epsfig{file=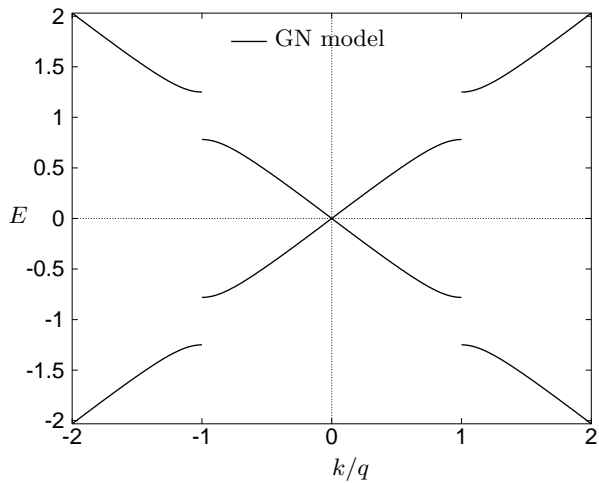, width=6cm, angle=-90}
    \end{center}
  \end{psfrags}
  \caption{Same as  Fig. 12, but for the discrete chiral GN model. The gaps are due to the periodic scalar potential.}
\end{figure}

Now consider the discrete chiral GN model at $T=0$. Here the potential is purely scalar and 
identical for matter and antimatter. The single particle spectrum is sketched in Fig. (14) for the high density case
($S_{\pm 1}\neq 0$ only), showing the appearance of a symmetric pair of gaps. (For lower densities, more
gaps would appear, but they are not relevant for the argument.) The filling of the single particle orbits
is indicated in Fig. (15) which now replaces Fig. (13). The common theme is evidently   
gap formation right at the Fermi surface. 

\begin{figure}[t]
  \begin{psfrags}
    \begin{center}
      \psfrag{MU0}{$\mu=0$}
      \psfrag{MUG0}{$\mu > 0$}
      \psfrag{MUK0}{$\mu < 0$}
      \epsfig{file=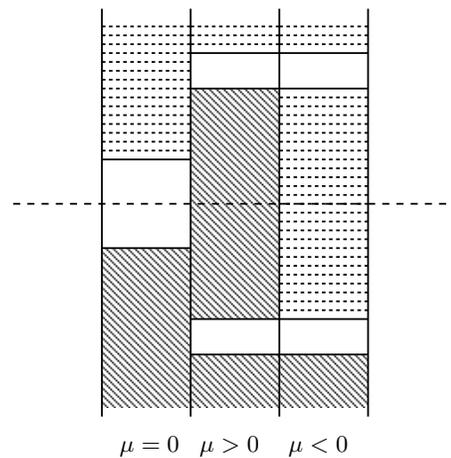, width=6cm}
    \end{center}
  \end{psfrags}
  \caption{Same as Fig. 13, but for the discrete chiral GN model. As in Fig. 13, there is always a  gap at the Fermi surface.}
\end{figure}

Turning to non-zero temperature the phase diagrams of the two GN models are very different,
both from each other and from what had previously been thought. The continuous chiral model possesses
two phases, the ``chiral spiral" and the chirally symmetric massless Fermi gas, separated by a horizontal
2nd order line $T=T_c$  \cite{R2}. Temperature affects only the radius of the helix, chemical potential
affects only the helix angle, and the two act independently. By contrast the discrete chiral model exhibits three 
distinct phases: Massless and massive Fermi gas and kink-antikink crystal. We have not been able to compute 
the full phase diagram analytically in this case but had to resort to a numerical solution of the Dirac-Hartree-Fock
equations. The phase boundary of the chirally restored phase and the grand canonical potential at large $\mu, \beta$
can be understood perturbatively. We find that the tricritical point of the old solution is replaced by a
Lifschitz point separating the homogeneous from the periodic ordered phase. 
Due to thermal effects which smear out the Fermi surface, the massive Fermi gas phase survives at higher
temperatures and small $\mu$. 

We have not found any first order transition in our calculations. As suggested by Fig. (9) the crystal phase takes
over the role played by the mixed phase in the old solution, although the quantitative details are quite different.
As explained in Ref. \cite{R5}, droplets of chirally restored matter with extra fermions can be interpreted as
``bag model" baryons with a mass of $M_B= N/\sqrt{2}$. This ``wrong" baryon mass leaves its traces in the old
phase diagram of the GN model, for instance in the value of $\mu_c$ where the first order phase
transition takes place at $T=0$ (the droplets fill all space), or equivalently in the slope of the energy density ${\cal E}_{HF}
(\rho)$ at $\rho=0$. In the revised phase diagram for the discrete chiral model, the correct, lower baryon mass
($M_B=2 N/\pi$) appears
instead. The continuous chiral model has massless baryons and consequently the structure of
the phase diagram changes qualitatively. Again, the behaviour of ${\cal E}_{HF} (\rho)$ at $\rho=0$ (vanishing slope)
reflects the presence of massless baryons and breakdown of translational invariance is shifted to the point $\mu=0$.
This shows that we have achieved an overall picture of the phase structure of both GN models which is now
consistent with their different baryon spectra and therefore physically more reasonable.

\begin{acknowledgements}
We thank Joshua Feinberg for a helpful correspondence.
\end{acknowledgements}

\begin{center}
{\bf APPENDIX:}
\end{center}
{\bf  Asymptotics at large $\mu$ and $\beta$}
\vskip 0.2cm
At large $\mu$ and $\beta$ one can use perturbation theory to determine $S_{\pm 1}$.
Starting point is the grand canonical potential  of Eq. (\ref{t6}). Single particle energies can be
computed in ADPT, Eq. (\ref{d11}).
Denoting $\tilde{S}_1={\rm i}S_1$ by $s$ and $q_1=\pi/a$ by $q$ to ease the notation, find
\begin{eqnarray}
\frac{\Psi}{N} & = & -\frac{2}{\beta}
\int_0^{q/2} \frac{{\rm d}k}{2\pi} \ln \left[ (1+{\rm e}^{-\beta(E_1-\mu)})
(1+{\rm e}^{\beta(E_1+\mu)})\right] \nonumber \\
& & -\frac{2}{\beta}
\int_{q/2}^{q} \frac{{\rm d}k}{2\pi} \ln \left[ (1+{\rm e}^{-\beta(E_2-\mu)})
(1+{\rm e}^{\beta(E_2+\mu)})\right] \nonumber \\
& & -\frac{2}{\beta}
\int_{q}^{3q/2} \frac{{\rm d}k}{2\pi} \ln \left[ (1+{\rm e}^{-\beta(E_3-\mu)})
(1+{\rm e}^{\beta(E_3+\mu)})\right] \nonumber \\
& & -\frac{2}{\beta}
\int_{3q/2}^{\Lambda} \frac{{\rm d}k}{2\pi} \ln \left[ (1+{\rm e}^{-\beta(E_1-\mu)})
(1+{\rm e}^{\beta(E_1+\mu)})\right] \nonumber \\
&&+ \frac{\Lambda^2}{2\pi} + \frac{\Lambda \mu}{\pi} + \frac{1}{\pi}  \ln (2\Lambda) s^2   \ ,
\label{t22}
\end{eqnarray}
with
\begin{eqnarray}
E_1 &=& k+\frac{s^2}{2}\left( \frac{1}{k+q}+ \frac{1}{k-q}\right) \ , \nonumber \\
E_2 &=& q - \sqrt{(k-q)^2+s^2}+ \frac{s^2}{2(k+q)}\ , \nonumber \\
E_3 &=& q + \sqrt{(k-q)^2+s^2}+ \frac{s^2}{2(k+q)}\ . 
\label{t23}
\end{eqnarray}
Asymptotically, we may use $\mu = - q$, cf. Fig. (7). Then, integrals involving
\begin{equation}
{\rm e}^{-\beta(E_1-\mu)} \ , \ \ {\rm e}^{-\beta(E_2-\mu)} \ , \ \ {\rm e}^{-\beta(E_3-\mu)}
\label{t24}
\end{equation}
are exponentially suppressed and can be neglected. Integrals involving 
\begin{equation}
{\rm e}^{\beta(E_1+\mu)}   
\label{t25}
\end{equation}
are negligible in the first term of Eq. (\ref{t22}) ($k=0...q/2$), while in the fourth term 
($k=3 q/2...\Lambda$) one has to pull out the $T=0$ piece as follows,
\begin{equation}
1+ {\rm e}^{\beta(E_1+\mu)} = {\rm e}^{\beta (E_1+\mu)} \left( 1+{\rm e}^{-\beta (E_1+\mu)}\right) \ .
\label{t26}
\end{equation}
Now, the $T \neq 0$ part is negligible. Finally, in the 3rd term of Eq. (\ref{t22}), we again
decompose
\begin{equation}
1+ {\rm e}^{\beta(E_3+\mu)} = {\rm e}^{\beta (E_3+\mu)} \left( 1+{\rm e}^{-\beta (E_3+\mu)}\right) \ ,
\label{t26a}
\end{equation}
but have to keep both pieces. In this way we find
\begin{eqnarray}
\frac{\Psi}{N} & = & -\frac{2}{\beta}
\int_{q/2}^{q} \frac{{\rm d}k}{2\pi} \ln (1+{\rm e}^{\beta(E_2+\mu)}) 
\nonumber \\
& & -\frac{2}{\beta}
\int_{q}^{3q/2} \frac{{\rm d}k}{2\pi} \ln (1+{\rm e}^{-\beta(E_3+\mu)}) \nonumber \\
& & -2 \int_{q}^{3 q/2} \frac{{\rm d}k}{2\pi} (E_3+\mu) -2 \int_{3 q/2}^{\Lambda} \frac{{\rm d}k}{2\pi} (E_1+\mu) 
\nonumber \\
& & + \frac{\Lambda^2}{2\pi}
+ \frac{\Lambda \mu}{\pi} + \frac{1}{\pi}  \ln (2\Lambda) s^2   \ . 
\label{t27}
\end{eqnarray}
Finally, the following approximations can be made: Evaluate the integrals over $E_1$ and $E_3$ exactly and expand
the result for small $s$ up to second order, keeping the logarithmic corrections. In the other integrals, drop the
$s^2$-term in $E_2,E_3$ and extend the integration limits to ($-\infty...q$) and ($q...\infty$), respectively.
Keeping only the $s$-dependent terms in $\Psi$ yields the simple final result
\begin{eqnarray}
\frac{\Psi}{N}  &=& -\frac{s^2}{4\pi}\left[1- 2\ln (4 s q)\right]
\nonumber \\
& & -\frac{2}{\pi \beta} \int_0^{\infty} {\rm d}k \ln \left( 1+{\rm e}^{-\beta \sqrt{k^2+s^2}}\right) \ .
\label{t28}
\end{eqnarray}
All of these approximations have been checked numerically against the full result.

Before proceeding, let us compare this result with the one which assumes unbroken
translational invariance, but taken at $T\neq 0,\mu=0$,
\begin{eqnarray}
\frac{\Psi}{N} &=& - \frac{m^2}{4\pi}(1-\ln m^2)
\nonumber \\
&  & -\frac{2}{\pi \beta}
\int_0^{\infty} {\rm d}k \ln \left( 1+ {\rm e}^{-\beta \sqrt{m^2+k^2}}\right) \ .
\label{t29}
\end{eqnarray}
The structure is almost the same, since $\mu$ has been ``eaten up" by $q$ from the single particle energies
in Eq. (\ref{t28}). Let us recall how to evaluate $m(T)$ at $\mu=0$ (here, the old solution was correct).
Take the derivative of the grand canonical potential with respect to $m$ and set it equal to zero,
\begin{equation}
m \left[ \ln (m) +2 \int_0^{\infty}\frac{{\rm d}k}{\sqrt{m^2+k^2}}
\left( \frac{1}{{\rm e}^{\beta \sqrt{m^2+k^2}}+1}\right) \right] = 0 \ .
\label{t30}
\end{equation}
Rescaling the integration variable and setting $\tilde{m}=\beta m$ then yields the 
solution
\begin{equation}
\beta = \tilde{m} \exp \left\{ 2 \int_0^{\infty} \frac{{\rm d}\kappa}{\sqrt{\tilde{m}^2+\kappa^2}}
\frac{1}{{\rm e}^{\sqrt{\tilde{m}^2+\kappa^2}}+1}\right\} \ .
\label{t31}
\end{equation}
If we evaluate the function on the r.h.s. and plot $\tilde{m}/\beta$ against $1/\beta$, we
recover the usual result for the temperature dependent mass, vanishing at $T_c={\rm e}^{\rm C}/\pi$.
That the Euler constant appears in $T_c$ follows from the following useful relation \cite{R12},
\begin{equation}\int_0^{\infty} \frac{{\rm d}q}{\sqrt{a^2+q^2}}
\frac{1}{{\rm e}^{\sqrt{a^2+q^2}}+1}
 = -\frac{1}{2}\ln \frac{a}{\pi}-\frac{1}{2} {\rm C}
+ O(a^2) \ .
\label{t32}
\end{equation}
If we now go back to Eq. (\ref{t28}) for the periodic case and minimize with respect to $s$, we find
\begin{equation}
s \left[ \ln (4s q) +2 \int_0^{\infty}\frac{{\rm d}k}{\sqrt{s^2+k^2}}
\left( \frac{1}{{\rm e}^{\beta \sqrt{s^2+k^2}}+1}\right) \right] = 0 \ .
\label{t33}
\end{equation}
This can be solved explicitly for $q$ as a function of $s,\beta$. The result can
of course only be trusted if $q$ and $\beta$ are large enough. We can reduce 
this problem to the preceding one (GN model at $\mu=0$)
as follows: The relation
between $4 q s$ and $4 q T$ in the crystal model is given by the same ``universal
curve" as the one between $m$ and $T$ at  $\mu=0$.

We finally note the following consequences of our asymptotic analysis ($\mu>0$):
The boundary line in the ($s,\mu$) plane at large $\mu$ and $T=0$ is given by
\begin{equation}
s=\frac{1}{4\mu}\ ,
\label{t34}
\end{equation}
in agreement with Eq. (\ref{d18}). The boundary line in the ($T,\mu$) plane at $s=0$
has the asymptotic form
\begin{equation}
T=\frac{{\rm e}^{\rm C}}{4\pi \mu}\ .
\label{t35}
\end{equation}

\vskip 0.5cm 
\

\end{document}